\documentclass{article}
\usepackage{setspace}
\usepackage[hidelinks]{hyperref}
\usepackage{amsmath}
\usepackage{parskip} 
\usepackage{natbib}
\usepackage{mathrsfs}
\usepackage{amsthm}

\usepackage{bm}
\usepackage{booktabs}
\usepackage{graphics}
\usepackage{multicol}
\usepackage[bf]{caption}
\usepackage{amssymb}
\usepackage{multirow}
\usepackage{booktabs}
\usepackage{subcaption}

\usepackage{comment}
\usepackage[margin=1in]{geometry}
\usepackage{floatrow}
\newfloatcommand{capbtabbox}{table}[][\FBwidth]
\usepackage[titletoc]{appendix}
\usepackage[graphicx]{realboxes}
\usepackage{gensymb}
\usepackage{color}

\usepackage{mathtools}

\title{Effective elasticity of a medium with many parallel fractures
}
\author{Filip P. Adamus\footnote{
Department of Earth Sciences, Memorial University of Newfoundland, Canada, {\tt adamusfp@gmail.com}}\,\,\,}

\date{}

\begin{document}
\maketitle
\begin{abstract}
We consider an alternative way of obtaining the effective elastic properties of a cracked medium.
Similarly, to the popular linear-slip model, we assume flat, parallel fractures, and long wavelengths. 
However, we do not treat fractures as weakness planes of displacement discontinuity.
In contrast to the classical models, we represent fractures by a thin layer embedded in the background medium.
In other words, we follow the Schoenberg-Douma matrix formalism for Backus averaging, but we relax the assumptions of infinite weakness and marginal thickness of a layer so that it does not correspond to the linear-slip plane.
To represent the properties of a fracture, we need a fourth order elasticity tensor and a thickness parameter.
The effective tensor becomes more complicated, but it may describe a higher concentration of parallel cracks more accurately.
Apart from the derivations of the effective elasticity tensors, we perform numerical experiments in which we compare the performance of our approach with a linear-slip model in the context of highly fractured media.
Our model becomes pertinent if filled-in or empty cracks occupy more than one percent of the effective medium.
\\ \\
{\bf{Keywords:}} Anisotropy, Effective, Elastic, Fractures.
\end{abstract}
\section{Introduction}
The influence of cracks on the elastic properties of a medium has been a topic of interest for numerous researchers.
There are various models used to describe the effective elasticity parameters of a fractured material. 
Some authors assume short wavelength compared to the cracked structure so that crack-pore microgeometry and the properties of a fluid are essential~\citep[e.g.,][]{OConnellBudiansky}.
Others often focus on long wavelengths that are more suitable for seismic frequencies~\citep[e.g.,][]{GarbinKnopoff}.
Further, models differ depending on the shape of cracks assumed.
If they are ellipsoidal~(\citet{Eshelby},~\citet{Nishizawa},~\citet{Hudson94}), the analysis usually becomes quite complicated~\citep{Hudson81}.
In practice, however, the aspect ratio of cracks is typically low.
Also, the details of their microstructure are often neglected in the seismic fracture-detection studies.
Therefore, cracks are not rarely described as flat~\citep[see][]{Kachanov}, which is a useful simplification, since in some cases the results do not change very much compared to the ellipsoidal shapes~(\citet{Hudson81}, \citet{SchDouma},~\citet{Thomsen95}).
Flat fractures may be planar~\citep{Sch80}, elliptical~\citep{Hudson80}, or irregular~\citep{GrechkaEtAl}.
Moreover, cracks can be distributed randomly~\citep{Hudson80}, can be aligned~\citep{Thomsen95} or parallel~\citep{SchDouma}.
In this paper, we consider long-wave, effective elasticity of a medium that corresponds to the background rock with parallel sets of flat fractures.
Due to long-wavelength assumption, our investigation is pertinent---but not limited---to seismic studies.

There are three widely investigated, effective models that assume long wavelength and flat fractures~\citep{Krebes}.
These are the linear-slip model, penny-shaped crack model, and the combined model.
Below, we shortly describe each of them.

The linear-slip stands for the fracture interface across which the traction vector is continuous, but the displacement is not~\citep{Sch80}.
The displacement discontinuity linearly depends on traction.
This relation is governed by the second-order tensor, which authors often refer to as the excess fracture compliance.
\citet{SchDouma} are first to use the linear-slip concept in modelling the effective elasticity.
Their work is based on~\citet{Backus} average, in which the aforementioned discontinuity corresponds to an infinitely weak and thin, horizontal layer.
The work of~\citet{SchDouma} was further developed by~\citet{SchSayers} that considered any orientation of linear-slip interfaces, not only the horizontal one.
Another, but penny-shaped crack model was proposed by~\citet{GarbinKnopoff} and then further developed by~\citet{Hudson80}.
They use scattering formalism, where circular cracks are treated as scatterers.
Cracks can be either aligned in one direction or randomly distributed.
The expressions of~\citet{GarbinKnopoff} are accurate to the first order in the concentration of cracks, whereas the expressions of~\citet{Hudson80} to the second order.
The second-order expressions correspond to the interactions between cracks that are not included in the linear-slip model. 
The penny-shaped model is complicated but accounts for the microstructure properties. 
The combined model is tantamount to the linear-slip one, but additionally relates the micro characteristics to the interface. 
Such a model was shown, for instance, by~\citet{Hudson96}.
The authors use scattering formalism and assume that circular cracks are aligned and parallel. 
This way, they obtain the excess fracture compliance related to cracks' properties.
Subsequently, this second-order tensor can be used in the linear-slip model~\citep{Hudson99}.

In this paper, we propose another long-wave model in which cracks are flat.
However, we assume a neither planar nor circular shape.
Herein, we treat fractures as sets of thin parallel layers.
We follow the approach of~\citet{SchDouma}, where they use the matrix formalism based on the Backus average.
As opposed to the aforementioned authors, we do not assume that layers corresponding to fractures are infinitely weak and thin.
In other words, we abandon the linear-slip description.
In this way, the properties of fractures are represented by fourth-order elasticity tensor and layer thickness, instead of excess fracture compliance only.
In the text, we refer to this method as the generalised Schoenberg-Douma approach or, simply, the generalised approach.
The linear slip model of~\citet{SchDouma} can be extended to viscoelastic~\citep{Visco} or poroelastic~\citep{Poro} media.
Analogously, the extension can be made to the generalised method.
However,  due to the complexity of expressions, we focus on the elastic effects only.
Thus, we assume that fractures are filled with solidified material.
The properties of the filling material affect the elasticity parameters of the crack.

The main advantage of the generalised approach over the linear-slip model is that a high concentration of cracks is explicitly taken into account. 
The relaxation of infinite weakness and marginal thickness of cracks allows the representation of the elastic properties of a medium with many parallel fractures or the background rock with harder inclusions. 
The main body of the paper is dedicated to the comparison between the two aforementioned approaches.
A heavily fractured medium was also considered in the combined models.
Therein, the high concentration of cracks is described by, for instance, crack density parameter.
In the rest part of the paper, we discuss the generalised approach and the combined models in the context of the effective elasticity of a medium with many parallel fractures.
\section{Generalised Schoenberg-Douma approach}
Elastic properties of parallel layers can be accurately approximated by the effective stiffness parameters of a homogeneous medium, assuming a sufficiently long wavelength.
To obtain these effective parameters, consider a well-known Voigt's representation of a fourth-order elasticity tensor of arbitrary anisotropy,
\begin{equation}\label{voigt}
\bm{C}_i=
\begin{bmatrix}
c_{11_i} & c_{12_i} & c_{13_i} & c_{14_i} & c_{15_i} & c_{16_i}\\
c_{12_i} & c_{22_i} & c_{23_i} & c_{24_i} & c_{25_i} & c_{26_i}\\
c_{13_i} & c_{23_i} & c_{33_i} & c_{34_i} & c_{35_i} & c_{36_i}\\
c_{14_i} & c_{24_i} & c_{34_i} & c_{44_i} & c_{45_i} & c_{46_i}\\
c_{15_i} & c_{25_i} & c_{35_i} & c_{45_i} & c_{55_i} & c_{56_i}\\
c_{16_i} & c_{26_i} & c_{36_i} & c_{46_i} & c_{56_i} & c_{66_i}\\
\end{bmatrix}
\,.
\end{equation}
Such a matrix describes the elastic properties of the $i$-th thin layer. 
The above parameters can also be represented by three matrices proposed by~\citet{HelbigSch},
\begin{equation}
\bm{M}_i=
\begin{bmatrix}
c_{11_i} & c_{12_i} & c_{16_i} \\
c_{12_i} & c_{22_i} & c_{26_i} \\
c_{16_i} & c_{26_i} & c_{66_i} \\
\end{bmatrix}
\,,\,\,\,
\bm{N}_i=
\begin{bmatrix}
c_{33_i} & c_{34_i} & c_{35_i} \\
c_{34_i} & c_{44_i} & c_{45_i} \\
c_{35_i} & c_{45_i} & c_{55_i} \\
\end{bmatrix}
\,,\,\,\,
\bm{P}_i=
\begin{bmatrix}
c_{13_i} & c_{14_i} & c_{15_i} \\
c_{23_i} & c_{24_i} & c_{25_i} \\
c_{36_i} & c_{46_i} & c_{56_i} \\
\end{bmatrix}
\,.
\end{equation}
These $3\times3$ matrices allow one to homogenise a stack of thin layers having arbitrary anisotropy, using process analogous to~\citet{Backus} average.
Assume that layers are horizontal, and the $x_3$-axis denotes depth.
The elasticity parameters of a homogenised, long-wave equivalent medium are
\begin{equation}\label{Ne}
\bm{N}_e=\overline{(\bm{N}_i^{-1})}^{-1}\,,
\end{equation}
\begin{equation}\label{Pe}
\bm{P}_e=\overline{(\bm{P}_i\,\bm{N}_i^{-1})}\,\overline{(\bm{N}_i^{-1})}^{-1}\,,
\end{equation}
\begin{equation}\label{Me}
\bm{M}_e=\overline{\bm{M}_i-\bm{P}_i\bm{N}_i^{-1}\bm{P}_i^T}+\overline{\bm{P}_i\bm{N}_i^{-1}}\,\overline{(\bm{N}_i^{-1})}^{-1}\,\overline{\bm{N}_i^{-1}\bm{P}_i^{T}}\,,
\end{equation}
where bar denotes the average and $^T$ stands for a transpose.
The average is weighted by the layer thickness.
The above derivations are identical to the ones of~\citet{HelbigSch},~\citet{SchDouma}, and~\citet{SchMuir}. 
For simplicity, throughout the paper, we assume density-scaled parameters.

\begin{figure}
\includegraphics[width=\textwidth]{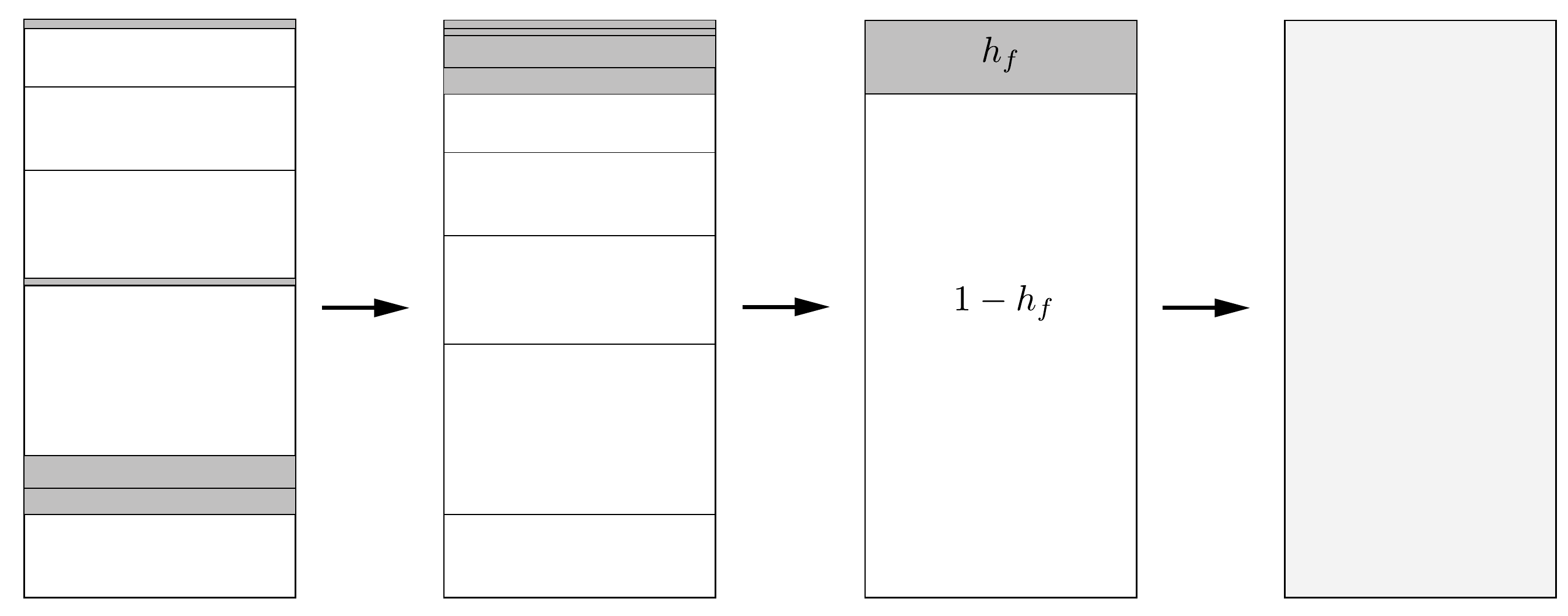}
\caption[\small{Commutative and associative properties of~\citet{HelbigSch} average}]{\small{The illustration of commutative and associative properties of~\citet{HelbigSch} average. The first column depicts the original layered medium, where grey colour denotes fractures filled with solidified material having different elastic properties. Subsequently, the layer sequence is interleaved so that fractures are cumulated in the upper part of the medium. Then, the effective parameters corresponding to fractures and background are  obtained, respectively. In the last column, the effective parameters for the homogenised medium are calculated. The intermediate steps have no influence on the final results but are useful in the evaluation of the fracture's effect.}}
\label{fig:one}
\end{figure}

We denote the relative thickness of a layer as $h_i$\,, where $i\in\{1,\dots,n\}$ and $\sum_{i=1}^nh_i=1$\,; thus, a medium is composed of numerous layers of various relative thicknesses. 
Some of these layers correspond to the background (host) medium, whereas the rest to the set of thin and long parallel fractures that are filled with a solidified material.
Since the average is commutative in the layer order and associative~\citep{SchMuir}, we can use these properties to fold the set of fractures into a single layer of total thickness $h_f$ and obtain its effective stiffnesses.
Analogously, we treat the background medium of total thickness $1-h_f$\,.
Below, we rewrite expressions~(\ref{Ne})--(\ref{Me}) in terms of background and fracture elasticities, indexed by letter $b$ and $f$, respectively.
\begin{equation}\label{Ne_new}
\bm{N}_e=\left((1-h_f)\bm{N}_b^{-1}+h_f\bm{N}_f^{-1}\right)^{-1}=\left((1-h_f)\bm{N}_b^{-1}+\bm{Z}\right)^{-1}\,,
\end{equation}
\begin{equation}
\bm{P}_e=\left((1-h_f)\bm{P}_b\bm{N}_b^{-1}+h_f\bm{P}_f\bm{N}_f^{-1}\right)\bm{N}_e\,,
\end{equation}
\begin{equation}\label{Me_new}
\begin{aligned}
\bm{M}_e&=(1-h_f)(\bm{M}_b-\bm{P}_b\bm{N}_b^{-1}\bm{P}_b^T)+h_f(\bm{M}_f-\bm{P}_f\bm{N}_f^{-1}\bm{P}_f^T)\\
&+\left((1-h_f)\bm{P}_b\bm{N}_b^{-1}+h_f\bm{P}_f\bm{N}_f^{-1}\right)\bm{N}_e\left((1-h_f)\bm{N}_b^{-1}\bm{P}_b^T+h_f\bm{N}_f^{-1}\bm{P}_f^T\right)\,,
\end{aligned}
\end{equation}
where $\bm{Z}$ is so-called fracture system compliance matrix (\citet{SchDouma},~\citet{SchSayers},~\citet{SchHelbig}).
We illustrate the homogenisation procedure used to obtain expressions~(\ref{Ne_new})--(\ref{Me_new}) in Figure~\ref{fig:one}.
Note that these expressions are the generalisations of~\citet{SchDouma} derivation. 
The aforementioned authors assumed that the  thickness of a system of fractures is marginal ($h_f\rightarrow0$) and that fractures are infinitely weak ($\bm{M}_f, \bm{N}_f, \bm{P}_f\rightarrow0$)\,. 
Upon introduction of such assumptions expressions~(\ref{Ne_new})--(\ref{Me_new}) reduce to their results, namely,
\begin{equation}\label{Ne_old}
\bm{N}_e\approx\left(\bm{N}_b^{-1}+h_f\bm{N}_f^{-1}\right)^{-1}=\left(\bm{N}_b^{-1}+\bm{Z}\right)^{-1}\,,
\end{equation}
\begin{equation}
\bm{P}_e\approx \bm{P}_b\bm{N}_b^{-1}\left(\bm{N}_b^{-1}+\bm{Z}\right)^{-1}\,,
\end{equation}
\begin{equation}\label{Me_old}
\bm{M}_e\approx \bm{M}_b-\bm{P}_b\bm{N}_b^{-1}\bm{P}_b^T+\bm{P}_b\bm{N}_b^{-1}\left(\bm{N}_b^{-1}+\bm{Z}\right)^{-1}\bm{N}_b^{-1}\bm{P}_b^T\,.
\end{equation}
Let us discuss the physical meaning of expressions~(\ref{Ne_old})--(\ref{Me_old}).
The effect of fractures is expressed by $\bm{Z}$ only, which stands for the excess compliance caused by total displacement discontinuity (total linear slip) across weakness planes~\citep{SchDouma}.
Thus, extremely thin layers are treated as planar discontinuities.
The average of a background medium with a set of horizontal weakness planes becomes a particular case of a more general theory of~\citet{SchSayers}, where planes of linear slip may have any orientation.
Specifically, consider an equation of~\citet{SchSayers} that describes a background medium with one set of parallel weakness planes,
\begin{equation}
s_{ijk\ell}=s_{ijk\ell_b}+s_{ijk\ell_f}=s_{ijk\ell_b}+\frac{1}{4}\left(Z_{ik}n_\ell n_j+Z_{jk}n_\ell n_i+Z_{i\ell}n_kn_j+Z_{j\ell}n_kn_i\right)\,,
\end{equation}
where $i,j,k,\ell\in\{1,2,3\}$\,, $s_{ijk\ell}$ denotes the compliances in a tensorial notation and $n_i$ indicates the orientation of the planar slip.
Note that if we insert vector $\bm{n}=[0,0,1]$\,, then we obtain the same result as from expressions~(\ref{Ne_old})--(\ref{Me_old}).
It is evident that in expressions~(\ref{Ne_new})--(\ref{Me_new}), neither marginal thickness nor infinite weakness of a layer corresponding to fractures is assumed.
Thus, expressions~(\ref{Ne_new})--(\ref{Me_new}) are the generalisations of~(\ref{Ne_old})--(\ref{Me_old}).
In this generalised approach, we do not follow the theory of linear-slip excess compliances presented by~\citet{SchSayers}.
We treat a set of parallel fractures as thin and weak layers that does not have to be infinitely thin and weak but are allowed to be so.
We believe that the aforementioned relaxation of linear-slip assumptions (no marginal thickness and infinite weakness) can be useful while willing to describe the effective elastic properties of a medium heavily cracked by weak fractures or a medium that contains few harder inclusions.

The physical meaning of the generalised approach can be extended to the influence of the set of parallel layers of any thickness and stiffness embedded in the background medium.
Note that it depends on more unknowns than Schoenberg-Douma approximation; thus, it becomes more complicated. 
The influence of the fractures (or set of layers of any stiffness) is governed by thickness $h_f$ and three matrices $\bm{M}_f$, $\bm{Z}$, and $\bm{P}_f$ (instead of $\bm{Z}$ only). 
Note that these three matrices represent a fourth-order elasticity tensor. 
\section{Examples of effective elasticity tensors}\label{tensors}
Let us consider quite a general example of a folded orthotropic layer of thickness $h_f$ embedded in an orthotropic background medium of thickness $h_b=1-h_f$\,.
We assume that tensors of both folded layer and background medium are expressed in a natural coordinate system.
The elasticity parameters of a layer are
\begin{gather}
\bm{M}_f=
\begin{bmatrix}
f_{11} & f_{12} & 0 \\
f_{12} & f_{22} &0 \\
0 & 0 & f_{66} \\
\end{bmatrix}
\,,\,\,\,
\bm{P}_f=
\begin{bmatrix}
f_{13} & 0 & 0 \\
f_{23} & 0 & 0 \\
0 & 0 & 0 \\
\end{bmatrix}
\,,\\ \nonumber \\
\label{NF}
\bm{N}_f=
\begin{bmatrix}
f_{33} & 0 & 0 \\
0 & f_{44} & 0 \\
0 & 0 & f_{55} \\
\end{bmatrix}
=
\begin{bmatrix}
h_fZ_{N}^{-1} & 0 & 0 \\
0 & h_fZ_{T_p}^{-1} & 0 \\
0 & 0 & h_fZ_{T_q}^{-1} \\
\end{bmatrix}
\,,\,\,\,
\end{gather}
where $f_{ij}$ stand for stiffnesses of a folded layer representing parallel fractures.
Subscript $_N$ denotes normal fracture system compliance, whereas $_{T_p}$ and $_{T_q}$ tangential compliances that, for horizontal layers, correspond to the $x_2$ and $x_1$ directions, respectively~\citep[see,][]{SchDouma}.
We assume neither marginal thickness nor infinite weakness of layers. 
To define the thickness of the folded layer, we use parameter $h_f$\,.
Now, we need to introduce a new parameter that could refer to the relative weakness of the embedded layer.
We propose
\begin{equation}
w_{ij}\equiv1-\frac{f_{ij}}{c_{ij_b}}\,,\qquad i,j\in\{1,\dots,6\}\,,
\end{equation}
where $c_{ij_b}$ are stiffnesses of a background medium. 
Weakness $w_{ij}$ is positive when the folded layer's elastic properties are weaker than the background, and negative when they are larger (we do not count unusual cases of negative stiffnesses).
Infinitely weak layer (meaning that its stiffnesses are close to zero) gives $w_{ij}\rightarrow1$\,.
Note that if all $w_{ij}=0$\,, then there is no distinction between background and folded layer.
A stiffness tensor describing the elastic properties of a background medium with a set of parallel layers is
\begin{equation}\label{matrix}
\bm{C}^{\rm{eff}}=
\begin{bmatrix}
\bm{c}_1 & 0\\
0 & \bm{c}_2 \\
\end{bmatrix}\,,
\end{equation}
where
\begin{gather}
\bm{c}_1=
\begin{bmatrix}
\vrule width 0pt height 20pt
\renewcommand*{\arraystretch}{1.2}
c_{11_b}\left(1-h_fw_{11}-h_b\frac{c_{13_b}^2}{c_{11_b}c_{33_b}}w_{13}\hat{\delta}_{N}\right) & c_{12_b}\left(1-h_fw_{12}-h_b\frac{c_{13_b}c_{23_b}}{c_{12_b}c_{33_b}}w_{13}w_{23}\hat{\delta}_{N}\right) &\,\,\,c_{13_b}(1-w_{13}\hat{\delta}_{N})\\
c_{12_b}\left(1-h_fw_{12}-h_b\frac{c_{13_b}c_{23_b}}{c_{12_b}c_{33_b}}w_{13}w_{23}\hat{\delta}_{N}\right) & c_{22_b}\left(1-h_fw_{22}-h_b\frac{c_{23_b}^2}{c_{22_b}c_{33_b}}w_{23}\hat{\delta}_{N}\right) &\,\,\, c_{23_b}(1-w_{23}\hat{\delta}_N) \\
c_{13_b}(1-w_{13}\hat{\delta}_{N}) & c_{23_b}(1-w_{23}\hat{\delta}_N) &\,\,\, c_{33_b}(1-w_{33}\hat{\delta}_{N}) 
\vrule width 0pt depth 10pt
\end{bmatrix}
 \raisetag{+0.05\baselineskip}
\end{gather}
and
\begin{equation}
\bm{c}_2=
\begin{bmatrix}
c_{44_b}(1-w_{44}\hat{\delta}_{T_p})  & 0 & 0\\ 
0 & c_{55_b}(1-w_{55}\hat{\delta}_{T_q}) & 0\\
0 & 0 & c_{66_b}\left(1-h_fw_{66}\right)\\
\end{bmatrix}\,.
\end{equation}
We define
\begin{equation}
0\leq\hat{\delta}_{N}\equiv \frac{Z_Nc_{33_b}}{1+Z_Nc_{33_b}-h_f}\leq1\,,
\end{equation}
\begin{equation}
0\leq\hat{\delta}_{T_p}\equiv \frac{Z_{T_p}c_{44_b}}{1+Z_{T_p}c_{44_b}-h_f}\leq1\,,
\end{equation}
\begin{equation}
0\leq\hat{\delta}_{T_q}\equiv \frac{Z_{T_q}c_{55_b}}{1+Z_{T_q}c_{55_b}-h_f}\leq1\,.
\end{equation}
Coefficients $\hat{\delta}_{N}$\,, $\hat{\delta}_{T_p}$\,, and $\hat{\delta}_{T_q}$ are similar to deltas shown in~\citet{SchHelbig}.
The essential difference is the presence of $h_f$ in our expressions, which makes them more general.
To indicate the above, we use hats over our parameters.
If $h_f\rightarrow 0$ and $w_{ij}\rightarrow1$\,, then matrix~(\ref{matrix}) represents the effective elasticity based on linear-slip theory.
If we only assume the infinite weakness of folded layer, meaning that $h_f\not\rightarrow 0$ and $w_{ij}\rightarrow1$\,, then the effective stiffnesses become weaker as compared to the stiffnesses based on linear-slip assumptions.
For instance, $c^{\rm{eff}}_{66}=c_{66_b}(1-h_f)$\,, whereas for linear-slip, $c^{\rm{eff}}_{66}=c_{66_b}$\,; it means that greater thickness of the folded layer, $h_f$\,, is responsible for the weakening of the effective medium.
Note that to describe the infinitely weak folded layer that corresponds to thick cavity or very soft inclusion, we need only four parameters: $Z_N$\,, $Z_{T_p}$\,, $Z_{T_q}$\,, and $h_f$ (see Appendix~\ref{ap:ch7_one}).
On the other hand, if we set $h_f\rightarrow 0$ and $w_{ij}\not\rightarrow1$\,, than the relaxed infinite weakness of the folded layer makes the effective medium stronger.

So far, we have discussed an example of an effective tensor corresponding to horizontal fractures embedded in a background medium. 
What if parallel fractures are not horizontal, but have a different orientation? 
What if there are more sets of fractures?
We propose to follow the recipe presented in the last section of~\citet{SchMuir}.
To model first set of fractures, we rotate the background medium to a desired coordinate system, then we calculate the effective parameters and rotate this tensor back. 
We repeat the process for other sets of fractures, where the background is the previously obtained effective medium.
The interaction between fractures is neglected.
Following the procedure of~\citet{SchMuir}, we obtain the effective tensor that corresponds to the orthotropic background medium with a set of orthotropic layers normal to the $x_1$-axis, namely,
\begin{gather}\label{mata1}
{\footnotesize{\bm{c}_1^{(1)}=}}
\left[
\begin{smallmatrix}
\vrule width 0pt height 15pt
c_{11_b}\left(1-w^{33}_{11}\hat{\delta}^{(1)}_{N}\right) \,\,\,& c_{12_b}\left(1-w^{23}_{12}\hat{\delta}^{(1)}_N\right)  &c_{13_b}\left(1-w_{13}\hat{\delta}^{(1)}_{N}\right)\\
c_{12_b}\left(1-w^{23}_{12}\hat{\delta}^{(1)}_N\right) \,\,\,& c_{22_b}\left(1-h_fw_{22}-h_b\frac{c_{12_b}^2}{c_{22_b}c_{11_b}}w^{23}_{12}\hat{\delta}^{(1)}_{N}\right) & c_{23_b}\left(1-h_fw^{12}_{23}-h_b\frac{c_{13_b}c_{12_b}}{c_{23_b}c_{33_b}}w_{13}w^{23}_{12}\hat{\delta}^{(1)}_{N}\right) \\
c_{13_b}\left(1-w_{13}\hat{\delta}^{(1)}_{N}\right) \,\,\,& c_{23_b}\left(1-h_fw^{12}_{23}-h_b\frac{c_{13_b}c_{12_b}}{c_{23_b}c_{33_b}}w_{13}w^{23}_{12}\hat{\delta}^{(1)}_{N}\right) & c_{33_b}\left(1-h_fw^{11}_{33}-h_b\frac{c_{13_b}^2}{c_{11_b}c_{33_b}}w_{13}\hat{\delta}^{(1)}_{N}\right) 
\vrule width 0pt depth 15pt
\end{smallmatrix}
\right]
 \raisetag{+0.03\baselineskip}
\end{gather}
and
\begin{equation}\label{mata2}
\bm{c}_2^{(1)}=
\begin{bmatrix}
c_{44_b}\left(1-h_fw^{66}_{44}\right) & 0 & 0\\ 
0 & c_{55_b}\left(1-w_{55}\hat{\delta}^{(1)}_{T_q}\right) & 0\\
0 & 0 & c_{66_b}\left(1-w^{44}_{66}\hat{\delta}^{(1)}_{T_p}\right)\\
\end{bmatrix}\,,
\end{equation}
where
\begin{equation}
w^{ij}_{k\ell}=1-\frac{f_{ij}}{c_{k\ell_b}}\,,\qquad {\rm{for}}  \quad (i,j)\neq(k,\ell),\quad {\rm{where}} \quad  i,j,k,\ell\in\{1,\dots,6\}\,
\end{equation}
and
\begin{equation}
0\leq\hat{\delta}^{(1)}_{N}\equiv \frac{Z_Nc_{11_b}}{1+Z_Nc_{11_b}-h_f}\leq1\,,
\end{equation}
\begin{equation}
0\leq\hat{\delta}^{(1)}_{T_p}\equiv \frac{Z_{T_p}c_{66_b}}{1+Z_{T_p}c_{66_b}-h_f}\leq1\,,
\end{equation}
\begin{equation}
0\leq\hat{\delta}^{(1)}_{T_q}\equiv \frac{Z_{T_q}c_{55_b}}{1+Z_{T_q}c_{55_b}-h_f}\leq1\,.
\end{equation}
Herein, subscripts $_{T_p}$ and $_{T_q}$ correspond to tangential compliances in horizontal ($x_2$) and vertical ($x_3$) directions, respectively.
\citet{SchHelbig} denote them as $_{T_p}=\,_H$ and $_{T_q}=\,_{V}$\,.
Superscript $^{(1)}$ indicates that the $x_1$-axis is normal to the set of embedded layers.

If fractures are normal to the $x_2$-axis, then we get
\begin{gather}
{\footnotesize{\bm{c}_1^{(2)}=}}
\left[
\begin{smallmatrix}
\vrule width 0pt height 18pt
c_{11_b}\left(1-h_fw_{11}-h_b\frac{c_{12_b}^2}{c_{11_b}c_{22_b}}w^{13}_{12}\hat{\delta}^{(2)}_{N}\right) \,\,&
c_{12_b}\left(1-w_{12}^{13}\hat{\delta}^{(2)}_{N}\right) 
&
\,\,\,\,c_{13_b}\left(1-h_fw^{12}_{13}-h_b\frac{c_{12_b}c_{23_b}}{c_{13_b}c_{22_b}}w^{13}_{12}w_{23}\hat{\delta}^{(2)}_{N}\right) 
\\
c_{12_b}\left(1-w_{12}^{13}\hat{\delta}^{(2)}_{N}\right)  
&
c_{22_b}\left(1-w^{33}_{22}\hat{\delta}^{(2)}_{N}\right) 
& c_{23_b}\left(1-w_{23}\hat{\delta}^{(2)}_N\right) \\
c_{13_b}\left(1-h_fw^{12}_{13}-h_b\frac{c_{12_b}c_{23_b}}{c_{13_b}c_{22_b}}w^{13}_{12}w_{23}\hat{\delta}^{(2)}_{N}\right)\,\,
&
c_{23_b}\left(1-w_{23}\hat{\delta}^{(2)}_N\right) 
&
\,\,c_{33_b}\left(1-h_fw^{22}_{33}-h_b\frac{c_{23_b}^2}{c_{22_b}c_{33_b}}w_{23}\hat{\delta}^{(2)}_{N}\right) 
 \vrule width 0pt depth 12pt
\end{smallmatrix}
\right]
 \raisetag{+0.01\baselineskip}
\end{gather}
and
\begin{equation}
\bm{c}_2^{(2)}=
\begin{bmatrix}
c_{44_b}\left(1-w_{44}\hat{\delta}^{(2)}_{T_p}\right)  & 0 & 0\\ 
0 & c_{55_b}\left(1-h_fw^{66}_{55}\right) & 0\\
0 & 0 & c_{66_b}\left(1-w^{55}_{66}\hat{\delta}^{(2)}_{T_q}\right) 
 \vrule width 0pt depth 10pt
\end{bmatrix}\,,
\end{equation}
where
\begin{equation}
0\leq\hat{\delta}^{(2)}_{N}\equiv \frac{Z_Nc_{22_b}}{1+Z_Nc_{22_b}-h_f}\leq1\,,
\end{equation}
\begin{equation}
0\leq\hat{\delta}^{(2)}_{T_p}\equiv \frac{Z_{T_p}c_{44_b}}{1+Z_{T_p}c_{44_b}-h_f}\leq1\,,
\end{equation}
\begin{equation}
0\leq\hat{\delta}^{(2)}_{T_q}\equiv \frac{Z_{T_q}c_{66_b}}{1+Z_{T_q}c_{66_b}-h_f}\leq1\,.
\end{equation}
Herein, subscripts $_{T_p}$ and $_{T_q}$ correspond to tangential compliances in vertical ($x_3$) and horizontal ($x_1$) directions, respectively.
Superscript $^{(2)}$ indicates the normal to the set of embedded layers.

An example of effective tensor that corresponds to two sets of orthotropic layers normal to the $x_1$-axis and the $x_2$-axis that are embedded in the orthotropic background medium is complicated to present analytically.
One of possible ways to obtain such a tensor is to  treat coefficients of matrices $\bm{c}_1^{(1)}$ and $\bm{c}_2^{(1)}$ as background parameters and substitute them inside matrices $\bm{c}_1^{(2)}$ and $\bm{c}_2^{(2)}$\,.

All the examples discussed above can be easily reduced to cases of higher symmetry.
For instance, if the background medium and folded layer are transversely isotropic with the $x_3$ symmetry axis (VTI), then $c_{11_b}=c_{22_b}\,$, $c_{13_b}=c_{23_b}$\,, $c_{44_b}=c_{55_b}$\,, $c_{11_b}=c_{12_b}+c_{66_b}$\,, and $w_{11}=w_{22}$\,, $w_{13}=w_{23}$\,, $w_{44}=w_{55}$\,, $w_{11}=w_{12}+w_{66}$\,.
There are infinitely many examples of other effective tensors, which depend on the number of folded layers, their orientations and symmetry classes, and the symmetry class of the original background medium.   
These examples can be easily derived using expressions~(\ref{Ne_new})--(\ref{Me_new}) and rotations of the coordinate system.
\section{Numerical experiments}
Let us discuss what may be the influence of thickness and stiffnesses of the folded layer that are neglected in the effective elasticity tensor obtained using linear-slip assumptions.
To do so, we consider numerical experiments in which we focus on the relative error,
\begin{equation}\label{err}
err=\frac{||(\bm{C}_b-\bm{C}_l^{\rm{eff}})-(\bm{C}_b-\bm{C}^{\rm{eff}})||_2}{||\bm{C}_b-\bm{C}_l^{\rm{eff}}||_2}\times100\%=\frac{||\Delta_l -\Delta||_2 }{||\Delta_l ||_2}\times100\%\,,
\end{equation}
where subscript $_l$ indicates the linear-slip approximation, and $\bm{C}_b$ denotes the background elasticity tensor. 
In the error above, we try to understand the discrepancy between linear-slip and generalised approach in estimating the influence of fractures.
Therefore, to separate this influence from the background rock, we consider $\Delta_l$ not $\bm{C}^{\rm{eff}}_l$ in the denominator.
We assume that the values of the background matrix $\bm{C}_b$ are known. 
We use a VTI background stiffness matrix from~\citet{SchHelbig}, namely,
\begin{equation}\label{back}
\bm{C}_b=
\begin{bmatrix}
10 & 4 &2.5 & 0 & 0 & 0\\
4 & 10 & 2.5 & 0 & 0 & 0\\
2.5 & 2.5 & 6 & 0 & 0 & 0\\
0 & 0 & 0 & 2 & 0 & 0\\
0 & 0 & 0 & 0 & 2 & 0\\
0 & 0 & 0 & 0 & 0 & 3\\
\end{bmatrix}\,.
\end{equation}
To describe the influence of cracks, in Schoenberg-Douma approximation, we need the excess fracture compliance $3\times3$ matrix $\bm{Z}=h_f\bm{N}_f^{-1}$ only.
Hence, in general, we require the maximum number of six independent compliances or, equivalently, six independent stiffnesses, and one thickness parameter (both matrices are symmetric).
However, to obtain the generalised formulas, apart from $\bm{Z}$\,, we need $3\times3$ matrices $\bm{M}_f$\,, $\bm{P}_f$\,, and thickness $h_f$\, ($\bm{P}_f$ is not symmetric).
It gives the maximum number of twenty--one independent elasticity parameters (if the folded layer is generally anisotropic) and one thickness coefficient.
In the numerical experiments, we assume that values of $\bm{Z}$ are the same for both approaches.
In other words, $\bm{Z}$ does not influence $err$\,.

We assume one set of parallel fractures with a normal directed towards the $x_1$-axis.
Herein, to manipulate the overall elastic properties of the folded layer easily and to understand its influence on $err$ better, we also assume that the background and folded layer's stiffnesses are proportional. Hence, in our example, the fractures---same as the background---have VTI symmetry.
We introduce,
\begin{equation}\label{k}
\bm{C}_f=k\,\bm{C}_b\,,
\end{equation}
where $k$ is a scalar denoting hardness of the folded layer and $\bm{C}_f$ is a $6\times6$ matrix that consists of fracture stiffnesses $f_{ij}$ (previously described by matrices $\bm{N}_f=h_f\bm{Z}^{-1}$\,, $\bm{M}_f$\,, and $\bm{P}_f$\,).
Factor $k$ is helpful, since one parameter governs all twenty--one stiffnesses of $\bm{C}_f$\,.
Also, the simplicity of $k$ can be physically justified when the folded layer is weak, and the exact values of specific stiffnesses do not matter so much.
Hardness $k$ can be understood as a simplification and an alternative to the previously defined weaknesses $w_{ij}$\,, where $k=1-w_{ij}$\,.
In the context of the above expressions, the parameters needed for the fracture description in Scohenberg-Douma approximation are
\begin{equation}\label{vol}
\bm{Z}=h_f
\begin{bmatrix}
f_{33} & f_{34} & f_{35}\\
f_{34} & f_{44} & f_{45}\\
f_{35} & f_{45} & f_{55}\\
\end{bmatrix}^{-1}
=
\frac{h_f}{k}
\begin{bmatrix}
c_{33_b} & c_{34_b} & c_{35_b}\\
c_{34_b} & c_{44_b} & c_{45_b}\\
c_{35_b} & c_{45_b} & c_{55_b}\\
\end{bmatrix}^{-1}
=
\frac{h_f}{k}
\begin{bmatrix}
1/6 & 0 & 0\\
0 & 1/2 & 0\\
0 & 0 & 1/2\\
\end{bmatrix}\,.
\end{equation}
In the generalised formulation, we also have the same two unknowns that describe the fractures (see Appendix~\ref{ap:ch7_two}).
Hence, the error depends only on the thickness $h_f$ and hardness $k$\,. 
Below, we perform three numerical experiments in which we manipulate the values of $h_f$ and $k$\,, so that either one or two of the linear-slip assumptions are relaxed.
Specifically, we relax $k\rightarrow0$\,, then $h_f\rightarrow0$\,, and lastly, we relax them both.
We check what the influence of the aforementioned relaxations on the relative error~(\ref{err}) is.

Let us make a brief comment on the volatility of $\bm{Z}$\,.
As we see in expression~(\ref{vol}), $\bm{Z}$ depends on hardness and thickness of fractures.
In the inverse problems, it might be difficult to estimate its values precisely, especially when the layer is very thin and weak (linear-slip theory).
If, say $h_f=10^{-12}$\,, then it does not really matter---in terms of marginal differences in the absolute values---if $k=100h_f$ or $k=0.01h_f$\,,  still $k$ is very small, but its influence on $\bm{Z}$ is enormous.
Hence, if fractures are very thin and weak, a small change in their compliances makes $\bm{Z}$ almost impossible to estimate (if we know the elastic properties of the effective medium, but do not know the background).
Therefore, to make our experiments more realistic, we do not allow $h_f$ and $k$ to be smaller than $10^{-6}$\,. 
\subsubsection*{Relaxation of infinite weakness assumption}
\begin{figure}[!htbp]
\includegraphics[scale=0.55]{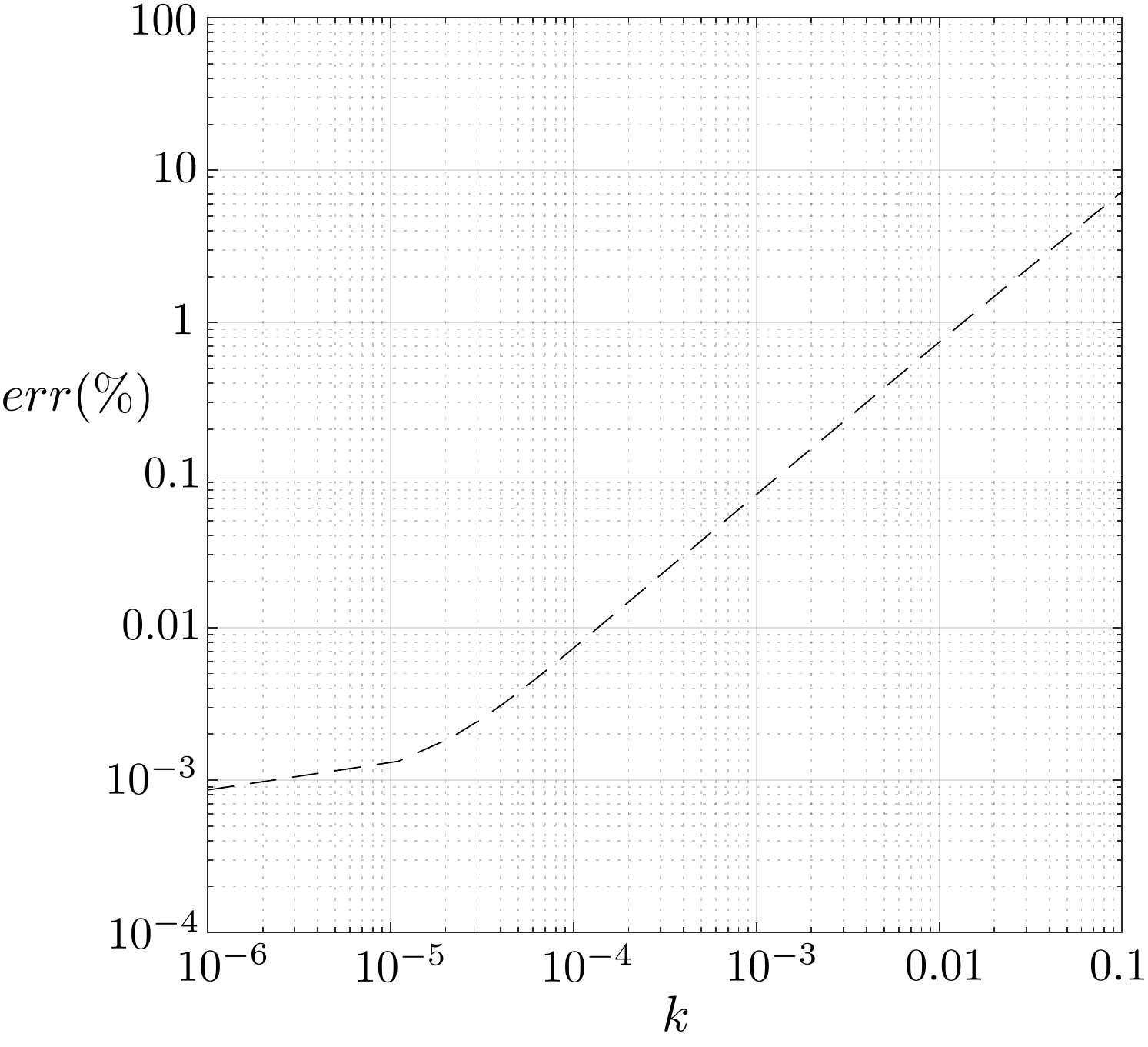}
\caption[\small{Relative error as a function of hardness, $k$\,, of folded layer}]{\small{Dashed line illustrates the relative error, $err$, as a function of hardness, $k$\,, of folded layer.
Thickness is fixed, $h_f=10^{-5}$\,; hence, values of $\bm{Z}$ diminish when $k$ grows. The axes are presented in a logarithmic scale.}}
\label{fig:two}
\end{figure}

In this experiment, we fix a very small thickness $h_f=10^{-5}$ and allow $k$ to grow.
Notice that when $k$ increases, $\bm{Z}$ becomes smaller.
Marginal $h_f$ and growing $k$ corresponds to the relaxation of the infinite weakness assumption of the linear-slip theory.
In this way, we wish to isolate the influence of the hardness of the folded layer on $err$.
Specifically, we check how much one can be wrong when in forward modelling assumes infinite weakness and marginal thickness of the folded layer, but the former assumption is incorrect.
The results are illustrated in Figure~\ref{fig:two}.

We see that the relaxation of the infinite weakness assumption has quite substantial effect on the results.
Let us think of extremely thin parallel inclusions that are ten times weaker than the background medium.
The above-mentioned physical example corresponds to $k=0.1$ for which $err$ is around seven percent.
Note that the error remains above one percent even for the inclusions fifty times weaker than the surroundings. 
Thus, despite the complexity of expressions~(\ref{Ne_new})--(\ref{Me_new}), the application of these generalised equations might be worth consideration if fractures are not extremely weak.
Matrix $\bm{Z}$ can have very low values if $k$ is much larger than $h_f$, which corresponds to the right part of Figure~\ref{fig:two}. 
\subsubsection*{Relaxation of marginal thickness assumption}
\begin{figure}
\includegraphics[scale=0.55]{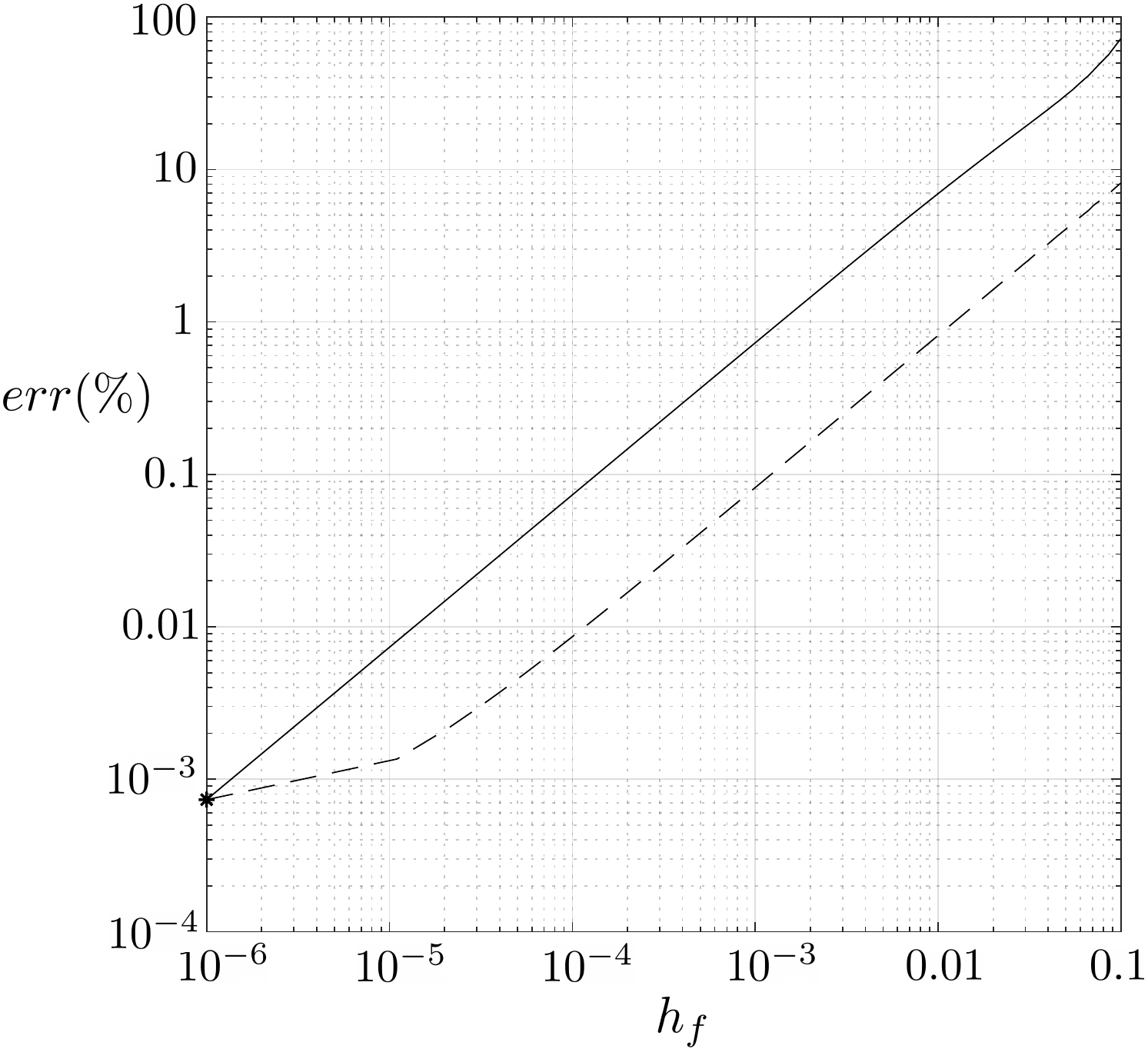}
\caption[\small{Relative error as a function of thickness, $h_f$\,, of folded layer}]{\small{Both lines illustrate the relative error, $err$\,, as a function of thickness, $h_f$\,, of folded layer.
For dashed line, hardness is fixed, $k=10^{-5}$\,; hence, values of $\bm{Z}$ increase along with growing $h_f$\,.
For solid line, $h_f=0.1k$\,; thus, values of $\bm{Z}$ are fixed. 
Both lines present identical values in a star point, since  $k=10^{-5}=10h_f$\,. 
As $h_f$ grows the discrepancy between two lines is larger, which is caused by the influence of $k$ (matrix $\bm{Z}$ has no influence on the error).
The axes are presented in a logarithmic scale.}}
\label{fig:three}
\end{figure}

Herein, we follow the infinite weakness assumption of the linear-slip theory.
Thus, we fix a very small value of $k=10^{-5}$\,.
However, we relax the assumption of marginal thickness; therefore, we allow $h_f$ to grow.
Notice that as thickness increases, so do values of matrix $\bm{Z}$\,. 
Physically, minimal value of $k$ and growing $h_f$ may correspond to empty cavities or very soft inclusions embedded in the host medium. 
In this numerical experiment, we expect to isolate the effect of relative thickness $h_f$ on $err$\,. 
Precisely, we examine how much one can be wrong when in forward modelling assumes the linear-slip deformation, but the assumption of marginal thickness is incorrect.
The results are depicted by a dashed line in Figure~\ref{fig:three}.

The influence of $h_f$ on the error seems to be quite significant and similar to the impact of $k$ (compare Figures~\ref{fig:two} and~\ref{fig:three}).
Let us think of parallel cavities that take one percent of the effective medium's space and which stiffnesses are extremely weak.
The aforementioned scenario corresponds to $h_f=0.01$ for which $err$ is almost one percent.
The error becomes even more substantial for greater thicknesses of the folded layer.
Again, the application of the generalised equations might be worth consideration if $h_f$ is substantial.
The situation of large $h_f$ and extremely weak layer corresponds to very substantial values of $\bm{Z}$\,.
%
\subsubsection*{Relaxation of both assumptions}
In this example, we choose specific values of $\bm{Z}$ so that $h_f$ and $k$ are both allowed to grow.
Hence, we relax both assumptions of linear-slip deformation.
We want realistic values of excess compliance matrix, similar to those of~\citet{SchHelbig}.
Therefore, we choose $k=10h_f$ and get
\begin{equation}
\bm{Z}=
\begin{bmatrix}
1/60 & 0 & 0\\
0 & 1/20 & 0\\
0 & 0 & 1/20\\
\end{bmatrix}\,.
\end{equation}
Having the above parameters set, we obtain $||\Delta_l||_2\approx1.8\, [\rm{km}/\rm{s^2}]$\,, which indicates that the effect of fractures is relatively moderate.
The solid line presents the cumulative influence of growing $h_f$ and $k$ on the error in Figure~\ref{fig:three}.

The result is similar to the previous numerical experiment, where $k$ varied, but $h_f$ was very small.
For example, if $h_f=0.01$ and $k=0.1$, then $err\approx6.95\%$\,. On the other hand, in Figure~\ref{fig:two}, $k=0.1$ corresponds to $h_f=10^{-5}$ and $err\approx7.21\%$\,. Further, if $h_f=10^{-4}$ and $k=0.01$, then $err\approx0.73\%$\,. In Figure~\ref{fig:two}, $k=0.01$ corresponds to $h_f=10^{-5}$ and $err\approx0.75\%$\,. 
From our numerical example, we deduce that $err$ does not augment if both assumptions, instead of one, are relaxed.

To sum up, in general, the larger the thickness or hardness of the layer of interest, the greater the error.
Based on our example, thickness $h_f$ and hardness $k$ seem to have similar contributions to $err$\,. We believe that the linear-slip theory is relatively accurate if fractures of the effective medium take less than one percent of its space and are at least a hundred times weaker than the background.
Otherwise, we recommend using the generalised approach.
The number of parameters used in our method can be greatly reduced by introducing scaling factor $k$, as presented in the numerical experiments and exemplified in Appendix~\ref{ap:ch7_two}.
\section{Comparison with other approaches}
So far, we have discussed the generalised approach in the context of the linear-slip theory only. 
In this section, we compare it to the models that take into account the micro properties, such as the concentration of cracks. 
First, let us consider the penny-shaped crack models proposed by~\citet{Hudson80} and~\citet{Hudson99}.
As we have already discussed in Section~6.1, these models were derived based on the scattering formalism.
The concentration of scatterers (cracks) is represented by the crack density parameter, $e$\,.
The intrinsic limitation of the scattering approach is that scatterers must be diluted~\citep{Keller}.
Hence, the parameter responsible for the concentration of cracks, $e$\,, cannot be large.
This is a significant drawback compared to the generalised Schoenberg-Douma model since $h_f$ has no limitation.
Hudson models are derived for isotropic background and involve second rank tensor $\bar{\bm{U}}$ that represents the elastic properties of fractures.
Following the works of Hudson, we consider isotropic background, cracks with normal towards the $x_3$-axis, and rotationally invariant $\bar{\bm{U}}$ and $\bm{Z}$ (meaning that $Z_{T_p}=Z_{T_q}=Z_T$\,).
The elasticity parameters of the linear-slip model are tantamount to the parameters shown in~\citet{Hudson80} and~\citet{Hudson99}.
Specifically, using the linear-slip model, we get  
\begin{gather}\label{back}
{\small{\bm{C}=}}
\setlength{\arraycolsep}{1.6pt}
\begin{bmatrix}
c_{11_b}\left(1-\frac{c_{12_b}^2}{c^2_{11_b}}\delta_{N}\right) & c_{12_b}\left(1-\frac{c_{12_b}}{c_{11_b}}\delta_{N}\right) &c_{12_b}(1-\delta_{N}) & 0 & 0 & 0\\
c_{12_b}\left(1-\frac{c_{12_b}}{c_{11_b}}\delta_{N}\right) & c_{11_b}\left(1-\frac{c_{12_b}^2}{c^2_{11_b}}\delta_{N}\right) & c_{12_b}(1-\delta_{N}) & 0 & 0 & 0\\
c_{12_b}(1-\delta_{N}) & c_{12_b}(1-\delta_{N}) & c_{11_b}(1-\delta_{N}) & 0 & 0 & 0\\
0 & 0 & 0 & c_{44_b}(1-\delta_{T}) & 0 & 0\\
0 & 0 & 0 & 0 & c_{44_b}(1-\delta_{T}) & 0\\
0 & 0 & 0 & 0 & 0 & c_{44_b}\\
\end{bmatrix},
\raisetag{-0.01\baselineskip}
\end{gather}
where $c_{11_b}=c_{12_b}+2c_{44_b}$ and
\begin{equation}
\delta_{N}=\frac{Z_Nc_{11_b}}{1+Z_Nc_{11_b}}\,,\quad
\delta_T=\frac{Z_{T_q}c_{44_b}}{1+Z_{T_q}c_{44_b}}\,.
\end{equation}
To obtain Hudson models, we insert either (see expressions~(51)--(54) of~\citet{Hudson80})
\begin{equation}
Z_N=\frac{\frac{c_{11_b}}{c_{44_b}}e\bar{U}_{33}+O(e^2)}{c_{11_b}\left(1-\frac{c_{11_b}}{c_{44_b}}e\bar{U}_{33}-O(e^2)\right)}\,,
\qquad
Z_T=\frac{e\bar{U}_{11}+O(e^2)}{c_{44_b}\left(1-e\bar{U}_{11}-O(e^2)\right)}\,,
\end{equation}
or (see expression~(8) of~\citet{Hudson99})
\begin{equation}
Z_N=\frac{e\bar{U}_{33}}{c_{44_b}}+\Theta(e^2)\,,
\qquad
Z_T=\frac{e\bar{U}_{11}}{c_{44_b}}+\Theta(e^2)\,,
\end{equation}
inside of $\bm{C}$\,.
Both $O(e^2)$ and $\Theta(e^2)$ are second-order terms in crack density, responsible for the crack interactions.
Hence, penny-shaped crack models, up to the first-order in $e$\,, can be treated as linear-slip models with parameters related to cracks' specific microstructure (we call them combined models).
Assuming that cracks are infinitely weak---by means of Eshelby theory---components $\bar{U}_{11}$ and $\bar{U}_{33}$ can be related to the background stiffnesses~(\citet{Eshelby}, \citet{Budiansky},~\citet{Hudson99}),
\begin{equation}\label{40}
\bar{U}_{11}=\frac{16c_{11_b}}{3\left(3c_{11_b}-2c_{44_b}\right)}\,,
\qquad
\bar{U}_{33}=\frac{4c_{11_b}}{3\left(c_{11_b}-c_{44_b}\right)}\,.
\end{equation}
As indicated by~\citet{SayersKachanov}, second terms in the models of Hudson are not sufficient to account for higher concentration of cracks. From $e>0.2$ they start to exhibit meaningless behaviour (the aforementioned limitation of the scattering approach).
Most of the combined approaches neglect the second-order terms.
Some of them do not assume interactions among cracks (non-interaction approximation), which is accurate for small (or, in some cases, moderate) concentrations of cracks only~\citep{KS}.
The other methods, such as the self-consistent, differential, or Mori-Tanaka schemes, tend to overestimate the impact of cracks on the effective stiffness~\citep{Kachanov}.
As shown by simulations of~\citet{Saenger}, the differential method seems to provide the best results for a high concentration of cracks.
In the aforementioned schemes, density parameter $e$ can be replaced by second and fourth-order tensors that cover all orientation distributions of cracks in a unified way~\citep{Kachanov}.
Below, for simplicity, we focus on $e$ only.

The upside of the combined models is their ability to relate micro properties of cracks to excess fracture compliance $\bm{Z}$\,. Also, under certain conditions, they allow expressing $\bm{Z}$ in terms of the background stiffnesses (expression~(\ref{40})).
However, we need to emphasise their main downside in the context of heavily cracked media. 
The combined models assume that cracks are flat, meaning that their aspect ratio is vanishingly small ($\alpha\rightarrow0$).
To relate micro properties to the linear-slip, the volume fraction occupied by cracks, $\phi_f$\,, must be also very small that is tantamount to $h_f\rightarrow0$ assumed by~\citet{SchDouma}. 
Crack density combines both $\alpha$ and $\phi_f$\,, namely,
\begin{equation}
e=\frac{3\phi_f}{4\pi\alpha}\,.
\end{equation}  
Hence, assuming that the aspect ratio is small (but not infinitely small), large number of $e$ implies significant value of $\phi_f$\,.
In turn, large $\phi_f$ is tantamount to a significant $h_f$ that violates the assumption underlying the linear-slip theory.
Therefore, the value of $e$ seems to be limited intrinsically.

The generalised method seems to be more adequate in describing media with many parallel fractures than combined approaches since a large concentration of cracks corresponds to large $h_f$ that does not violate the generalised method's assumptions.
Perhaps, it is possible to utilise both parameters $e$ and $h_f$ jointly. 
In a particular case of flat but infinitely weak fractures (with no marginal thickness of the folded layer), we may use the Eshelby theory to express $\bm{Z}$ in terms of background elasticities and density parameter as it is done in the combined approaches.
In other words, we conjecture that 
\begin{equation}\label{41}
Z_{N}=\frac{4c_{11_b}e}{3c_{44_b}\left(c_{11_b}-c_{44_b}\right)}\,,
\qquad
Z_{T}=\frac{16c_{11_b}e}{3c_{44_b}\left(3c_{11_b}-2c_{44_b}\right)}\,
\end{equation} 
can be inserted inside matrices obtained using the generalised method (where $w_{ij}=1$ and $h_f>0$)\,.
This conjecture needs to be verified by experimental studies.

To sum up, a higher concentration of cracks can be either described by a density parameter or by $h_f$, depending on whether the combined model or generalised method is used, respectively.
Both methods give different effective elasticity parameters.
For instance, if background is isotropic and cracks are aligned along the axis, $e$ influences five independent effective stiffnesses (matrix~(\ref{back})), whereas $h_f$ influences six stiffnesses (simplified matrix~(\ref{matrix})).
Moreover, $e$ is accurate for small numbers only, whereas $h_f$ has no limitations.
Perhaps it is possible to combine micro properties with a particular case of the generalised approach employing the Eshelby theory.
\section{Conclusions}
We have presented an alternative way of computing the effective elasticity tensor corresponding to a medium with parallel sets of fractures that are filled with a solidified material.
We have discussed a traditional Schoenberg-Douma method that is based on the linear-slip approximation. 
Further, we have shown a generalisation of their approach and examined if consideration of more complicated expressions might be useful in the context of the approximation accuracy.
The significant difference between the two aforementioned approaches is that the generalisation considers thickness and additional (to $\bm{Z}^{-1}$) elastic properties of the layer that corresponds to the system of parallel fractures.
We believe that no assumption of linear-slip deformation in the generalised expressions can be useful while describing the effective elastic properties of a medium that is heavily fractured or contains a few harder inclusions.

In case a material includes numerous empty cavities, our model simplifies, so that the additional elastic properties of the folded layer are not taken into account (see Appendix~\ref{ap:ch7_one}).
However, in such a case, our approach still differs from a traditional linear-slip method since the thickness parameter ($h_f$) is considered.
Another simplification to our model is possible if we assume that the scaled background stiffnesses describe the elasticity of fractures. This way, only two additional parameters---thickness $h_f$ and scaling factor $k$---are needed to consider the influence of parallel fractures (see Appendix~\ref{ap:ch7_two}). The linear-slip model can be simplified in a similar manner.

Numerical experiments have exposed that in forward problems, the consideration of parallel fractures intensity (equivalently, thickness $h_f$ of the folded layer) and its additional elasticity parameters might be essential.
We believe that also in the inverse problems, where we expect a heavily cracked medium, the generalised equations shown in this paper might be worth considering.
It seems that the linear-slip approximation is quite accurate if fractures of the effective medium take less than one percent of its space and are at least a hundred times weaker than the background. 
If the fractures take more space or are harder, we recommend using the generalised Schoenberg-Douma approach that does not neglect the intensity of inclusions.

Other possible methods that take into account the high concentration of cracks are the combined, penny-shaped crack models.
These approaches take into consideration the density and microstructure of cracks. 
The drawback of these methods is that they are limited intrinsically to the diluted concentration of cracks, and they are quite complicated. 
Also, their parameter responsible for the intensity of cracks ($e$) affects less number of the effective stiffnesses compared to the analogous parameter presented in the generalised approach ($h_f$)\,.
A combination of penny-shaped crack models with the generalised method seems possible. In this way, cracks are described by the background elasticities, density parameter and $h_f\,$.

Note that the generalised Schoenberg-Douma method is suitable for the computation of long-wave effective elasticity of any medium composed of parallel layers. 
Naturally, this approach is not limited to a very thin layer embedded in the background medium, which was the focus of this paper.
\section*{Acknowledgements}
We wish to acknowledge discussions with Michael A. Slawinski. Also, we thank Elena Patarini for the graphical support.
The research was done in the context of The Geomechanics Project partially supported by the Natural Sciences and Engineering Research Council of Canada, grant 202259. The author has no conflict of interest to declare.
\bibliography{bibliography}
\bibliographystyle{apa}
\appendix
\section{Effective elasticity with weakness assumption only}\label{ap:ch7_one}
Consider an effective tensor that corresponds to the orthotropic background medium with a set of orthotropic layers normal to the $x_1$-axis. 
Layers are folded into one medium representing fractures. 
If we assume infinite weakness of the folded layer, but not its marginal thickness, then matrices~(\ref{mata1}) and (\ref{mata2}) are simplified. We get 
\begin{equation}
\bm{C}^{\rm{eff}}=
\begin{bmatrix}
\bm{c}_1 & 0\\
0 & \bm{c}_2 \\
\end{bmatrix}\,,
\end{equation}
where
\begin{equation}
\bm{c}_1=
\begin{bmatrix}
c_{11_b}\left(1-\hat{\delta}_{N}\right) & c_{12_b}\left(1-\hat{\delta}_N\right)  &c_{13_b}\left(1-\hat{\delta}_{N}\right)\\
c_{12_b}\left(1-\hat{\delta}_N\right) & c_{22_b}h_b\left(1-\frac{c_{12_b}^2}{c_{22_b}c_{11_b}}\hat{\delta}_{N}\right) & c_{23_b}h_b\left(1-\frac{c_{13_b}c_{12_b}}{c_{23_b}c_{33_b}}\hat{\delta}_{N}\right) \\
c_{13_b}\left(1-\hat{\delta}_{N}\right) & c_{23_b}h_b\left(1-\frac{c_{13_b}c_{12_b}}{c_{23_b}c_{33_b}}\hat{\delta}_{N}\right) & c_{33_b}h_b\left(1-\frac{c_{13_b}^2}{c_{11_b}c_{33_b}}\hat{\delta}_{N}\right) 
 \vrule width 0pt depth 13pt
\end{bmatrix}
\,,
\end{equation}
\begin{equation}
\bm{c}_2=
\begin{bmatrix}
c_{44_b}h_b & 0 & 0\\ 
0 & c_{55_b}\left(1-\hat{\delta}_{V}\right) & 0\\
0 & 0 & c_{66_b}\left(1-\hat{\delta}_{H}\right)
 \vrule width 0pt depth 13pt
\end{bmatrix}\,,
\end{equation}
and
\begin{equation}
\hat{\delta}_{N}=\frac{Z_Nc_{11_b}}{h_b+Z_Nc_{11_b}}\,,\quad
\hat{\delta}_V=\frac{Z_Vc_{55_b}}{h_b+Z_{V}c_{55_b}}\,,
\quad
\hat{\delta}_H=\frac{Z_Hc_{66_b}}{h_b+Z_{H}c_{66_b}}\,.
\end{equation}
Excess fracture compliance that corresponds to displacement in normal, vertical, and horizontal direction is denoted by $Z_N$\,, $Z_V\,$, and $Z_H$\,, respectively.
The elastic properties of a background medium are described by $c_{ij_b}$\,, whereas $h_b$ stands for the relative thickness of such medium.  
The description of fractures needs only four parameters; $Z_N$\,, $Z_V$\,, $Z_H$\,, and $h_b$\,.
Thickness $h_b\in(0,1]$ is the only coefficient that distinguishes the above matrices from the linear-slip description.
If $h_b=1$ than we get effective elasticity consistent with theory of~\citet{SchDouma} or~\citet{SchSayers}.
\section{Effective elasticity with scaling factor $k$}\label{ap:ch7_two}
Again, consider an effective tensor that corresponds to the orthotropic background with an embedded set of orthotropic fractures normal to the $x_1$-axis. Assume that the elastic properties of folded  fractures are equal to the scaled background stiffnesses. In other words, we invoke expression~(\ref{k}), namely, $C_f=kC_b$\,, where $k$ is a scalar that relates $6\times6$ matrices describing fractures and the background, respectively. Due to the assumption above, the effective tensor represented by matrices~(\ref{mata1}) and (\ref{mata2}) requires a lower number of parameters. To show it, first, we rewrite the weaknesses
\begin{equation}\label{loko1}
w_{ij}=1-k\,,\qquad w^{ij}_{kl}=1-kc_{ij_b}/c_{kl_b}
\end{equation}
and the excess compliances
\begin{equation}\label{loko2}
Z_N=\frac{h_f}{c_{33_b}k}\,, \qquad Z_{T_p}=\frac{h_f}{c_{44_b}k}\,, \qquad Z_{T_q}=\frac{h_f}{c_{55_b}k}\,.
\end{equation}
Subsequently, we insert expressions~(\ref{loko1}) and (\ref{loko2}) into matrices~(\ref{mata1}) and (\ref{mata2}), to obtain
\begin{equation}\label{kakaka}
\bm{C}^{\rm{eff}}=
\begin{bmatrix}
{c}_{11} & c_{12} & c_{13} & 0 & 0 & 0\\
{c}_{12} & c_{22} & c_{23} & 0 & 0 & 0\\
{c}_{13} & c_{23} & c_{33} & 0 & 0 & 0\\
0 & 0 & 0 & c_{44} & 0 & 0\\
0 & 0 & 0 & 0 & c_{55} & 0\\
0 & 0 & 0 & 0& 0 & c_{66}\\
\end{bmatrix}\,,
\end{equation}
where
\begin{equation}
c_{11}=\frac{c_{11_b}c_{33_b}k}{c_{11_b}h_f+c_{33_b}k-c_{33_b}h_fk}\,,
\end{equation}
\begin{equation}
\begin{aligned}
c_{22}&=h_fk\left(c_{22_b}-\frac{c_{23_b}^2}{c_{33_b}}\right)+(1-h_f)\left(c_{22_b}-\frac{c_{12_b}^2}{c_{11_b}}\right)\\
&\hphantom{XXXXXXXXXXXXX}+\frac{c_{11_b}c_{33_b}k\left(\frac{c_{12_b}(1-h_f)}{c_{11_b}}+\frac{c_{23_b}h_f}{c_{33_b}}\right)^2}{c_{11_b}h_f+c_{33_b}k-c_{33_b}h_fk}\,,
\end{aligned}
\end{equation}
\begin{equation}
\begin{aligned}
c_{33}&=h_fk\left(c_{11_b}-\frac{c_{13_b}^2}{c_{33_b}}\right)+(1-h_f)\left(c_{33_b}-\frac{c_{13_b}^2}{c_{11_b}}\right)\\
&\hphantom{XXXXXXXXXXXXX}+\frac{c_{11_b}c_{33_b}k\left(\frac{c_{13_b}(1-h_f)}{c_{11_b}}+\frac{c_{13_b}h_f}{c_{33_b}}\right)^2}{c_{11_b}h_f+c_{33_b}k-c_{33_b}h_fk}\,,
\end{aligned}
\end{equation}
\begin{equation}
c_{12}=\frac{c_{12_b}c_{33_b}k+c_{11_b}c_{23_b}h_fk-c_{12_b}c_{33_b}h_fk}{c_{11_b}h_f+c_{33_b}k-c_{33_b}h_fk}\,,
\end{equation}
\begin{equation}
c_{13}=\frac{c_{13_b}k\left(c_{33_b}+c_{11_b}h_f-c_{33_b}h_f\right)}{c_{11_b}h_f+c_{33_b}k-c_{33_b}h_fk}\,,
\end{equation}
\begin{equation}
\begin{aligned}
c_{23}&=h_fk\left(c_{12_b}-\frac{c_{13_b}c_{23_b}}{c_{33_b}}\right)+(1-h_f)\left(c_{23_b}-\frac{c_{12_b}c_{13_b}}{c_{11_b}}\right)\\
&\hphantom{XXXXXXXXX}+\frac{c_{11_b}c_{33_b}k\left(\frac{c_{13_b}(1-h_f)}{c_{11_b}}+\frac{c_{13_b}h_f}{c_{33_b}}\right)\left(\frac{c_{12_b}(1-h_f)}{c_{11_b}}+\frac{c_{23_b}h_f}{c_{33_b}}\right)}{c_{11_b}h_f+c_{33_b}k-c_{33_b}h_fk}\,,
\end{aligned}
\end{equation}
\begin{equation}
c_{44}=c_{44_b}-c_{44_b}h_f+c_{66_b}h_fk\,,
\end{equation}
\begin{equation}
c_{55}=\frac{c_{55_b}k}{h_f+k-h_fk}\,,
\end{equation}
\begin{equation}
c_{66}=\frac{c_{66_b}c_{44_b}k}{c_{66_b}h_f+c_{44_b}k-c_{44_b}h_fk}\,.
\end{equation}
The effective elasticity matrix~(\ref{kakaka}) is described by the background stiffnesses $c_{ij_b}$ and only two additional parameters $k$ and $h_f$ that are responsible for the set of parallel fractures.

\end{document}